\documentclass{PoS}
\usepackage{macros_static,mathrsfs,amssymb,amsmath,amsfonts,epsfig,cite,graphics}

\def\LALPHA{\hbox{\epsfxsize=2.0 true cm
                           \epsfbox{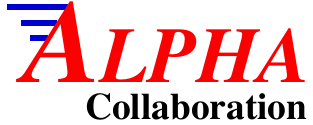}}
                   }

\title{$B$-meson physics from non-perturbative lattice heavy quark effective theory}

\ShortTitle{}

\author{
\hfill
\it Edinburgh 2011/06}

\author{
\speaker{Nicolas Garron } 
\textnormal{on behalf of the} \LALPHA  
\vspace{.5cm} \\
School of Physics and Astronomy, University of Edinburgh, Edinburgh EH9 3JZ, U.K.\\
}

\abstract{
During the last years, the ALPHA collaboration has been developing 
and implementing a method based on Heavy Quark Effective Theory 
(HQET) to compute B-mesons observables through lattice simulations. 
Thanks to a non-perturbative matching 
to QCD,
the theory is renormalizable at any order of the heavy quark mass expansion.
In order to extract precisely the relevant matrix elements and masses,
we use all-to-all propagators and solve an generalized eigenvalue problem 
(GEVP). We have shown in the quenched approximation that quantities
like the $b$-quark mass $\mb$, the heavy-light decay constant(s) 
or the $B$-meson spectrum can be computed precisely beyond the static
approximation (including the first corrections in $1/\mb$).
More recently, we have started to include the sea quark effects,
by working with $N_f=2$ light flavors of dynamical 
fermions~\cite{Blossier:2010vj}
.
The computation of the matching parameters is almost finished,
but concerning the extraction of the hadronic quantities 
for which we use some CLS ensembles \cite{CLS}, 
only one lattice spacing has been analyzed so far.
In this proceeding we report on the status of this project and 
present some preliminary results. The strategy is sketched  
in the following figure:
\begin{center}
\includegraphics*[width=9cm, height=7cm]{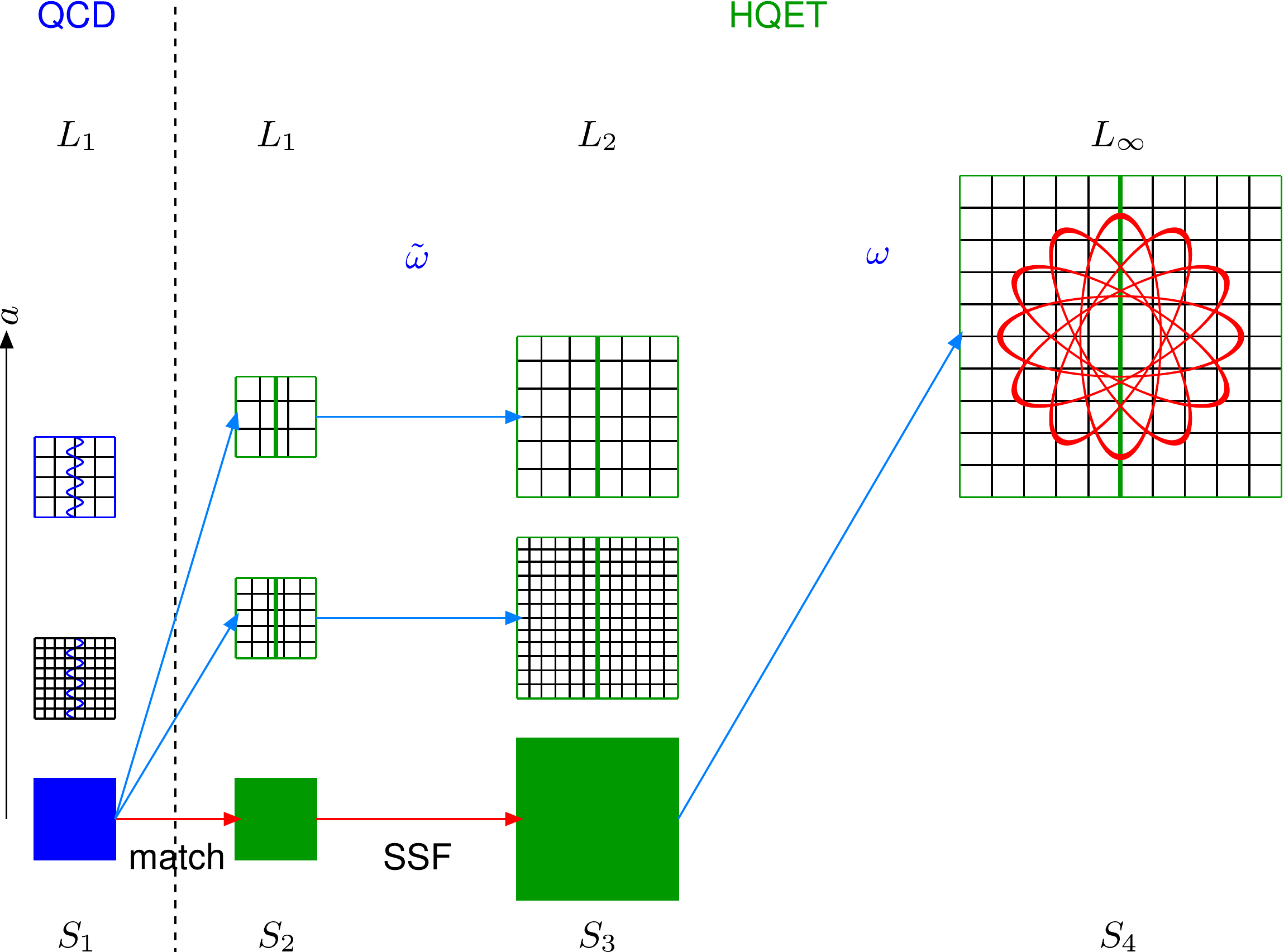}
\end{center}
}

\FullConference{35th International Conference of High Energy Physics - ICHEP2010,\\
		July 22-28, 2010\\
		Paris France}

\begin{document}

\section{Introduction}

The search for new physics is currently limited by the size of the theoretical 
uncertainties, mainly due to the errors affecting the hadronic quantities. 
This is specially true in the $B$ physics area, where in many cases the errors
are largely dominated by quantities like 
heavy-light decay constants or bag parameters,
which are obtained by lattice simulation. 
If the recent progress of the lattice community in the light quark sector 
are
very impressive, lattice simulations around the $b$-quark mass are still difficult.
Various strategies have been developed 
(see e.g.~\cite{Aubin:2009yh, DellaMorte:2010ym, Gamiz})
based on effective theories (NRQCD, HQET),
on relativistic formulation (Fermilab, RHQ) or a combination of both.
As already mentioned in the abstract, the strategy followed by the ALPHA collaboration 
is based 
on non-perturbative HQET in the static approximation 
and at the $1/\mb$ order
(see \cite{Sommer:2010ic} for a pedagogical introduction). 
This is a theoretically sound framework, in which the systematic 
errors are well under control.
Based on our experience in the quenched approximation, 
we believe that the total uncertainty affecting the results can be made small enough to have an impact 
in the search for new physics, and maybe unveil BSM effects.
This method has been tested in the quenched approximation to study 
the $B_{\rm s}$ meson spectrum \cite{Blossier:2010vz}, to determine 
the $b$-quark mass \cite{DellaMorte:2006cb} and the decay constant
$f_{\rm B_{\rm s}}$ \cite{Blossier:2010mk}. 
The computation is essentially done in two steps:
as explained in section 2, the HQET parameters are 
obtained from a non-perturbative matching to QCD in a small volume,
while the hadronic matrix element and masses are obtained from 
a subset of ensembles generated within the CLS effort (see section 3).
We present our conclusions together with our preliminary results in section 4.

\section{Computation of the relevant HQET parameters}

At the $1/m_{\rm b}$ order, we write the HQET Lagrangian 
in the following way:
\be
\lag{HQET}(x) =  \lag{stat}(x) - \omegakin\Okin(x)
        - \omegaspin\Ospin(x)  \,,
\ee
\bes 
\lag{stat}(x) = \heavyb(x) \,D_0\, \heavy(x) \;, \quad
\Okin(x) = \heavyb(x){\bf D}^2\heavy(x) \,,\quad
  \Ospin(x) = \heavyb(x){\boldsymbol\sigma}\!\cdot\!{\bf B}\heavy(x)\,.
\ees
We are also interested in the time component of the heavy-light axial current $A_0$.
Considering only the terms which contribute to zero spatial-momentum correlation functions,
we write it as
\bes
 \label{e:ahqet}
 \Ahqet(x)&=& \zahqet\,[\Astat(x)+  \cah{1}\Ah{1}(x)]\,, \\
 \Ah{1}(x) &=& \lightb(x){1\over2}
            \gamma_5\gamma_i(\nabsym{i}-\lnabsym{i}\;)\heavy(x)\,,
 \quad \Astat(x) = \lightb(x)\gamma_0\gamma_5\heavy(x)\,,
 \label{e:dahqet}
\ees
where $\nabsym{i}$ denotes the symmetric derivative.
The computation of $\omegakin, \omegaspin, \zahqet,  \cah{1}$ and $\mhbare$ 
is done following the strategy presented in~\cite{Blossier:2010jk,Blossier:2010vj}. 
In a volume of linear space extent $L_1\sim 0.5$ fm,
where the $b$-quark can be simulated with discretization effects under control,
we compute a set of observables $\Phi_{i=1,\ldots, 5}$
at four different lattice spacings and extrapolate them to the continuum
\footnote{In the continuum and large volume limits, $\Phi_1$ is proportional to 
the meson mass and $\Phi_2$ to the logarithm of the decay constant, respectively.
$\Phi_3$ is used to determine the counter-term of the axial current, 
and $\Phi_{4,5}$ for the determination of the kinetic and magnetic term,
respectively.}. 
In this set of simulations called $S_1$ in the figure, the light quark masses are set to 0, 
while for the (RGI) heavy quark mass $M$ we have chosen nine different values
covering a range between the charm to the bottom mass.
In another set of simulations called $S_2$, 
where we use the same value of the physical volume,
we compute the corresponding quantities 
in the effective theory (at the $1/\mb$ order) 
and match them to their QCD counterpart.
This matching can be written in the following way:
\begin{equation}
\label{eq:matchL1}
\Phi^{\rm QCD}_i(L_1,M,0) = \eta_i(L_1,a) + \sum_j \varphi_{ij}(L_1,a)\,\tilde\omega_j(M,a) \;,
\end{equation}
where $\eta$ and $\varphi$ are computed by lattice simulations 
for different values of the lattice spacing $a$. 
In other words the matching equations determine the set of parameters 
$\tilde\omega =  \varphi^{-1}\,[\Phi^{\rm QCD} -\eta]$.
We then compute $\eta$ and $\varphi$ in a larger volume of space extent $L_2=2L_1$,
and using the parameters $\tilde \omega(M,a)$ determined in the previous step
we compute the observables $\Phi(L_2,M,0)$ according to a formula similar to eq.~(\ref{eq:matchL1})
but where $L_1$ is replaced by $L_2$ and $\tilde\omega$ by $\omega$ 
(note that in this step called $S_3$
the continuum limit can be taken because the divergences
cancel out exactly).
This procedure can then be re-iterated until the volume reached is large enough 
for finite size effects to be negligible,
typically around $(2\, \fm)^3$. 
In practice, it turns out that three different volumes are enough 
($L_1,L_2$ and the large volume one). 
Thus, the HQET parameters that can be used in large volume simulations
(denoted as $S_4$) 
are given by
$
\omega(M,a) = \varphi^{-1}(L_2,a)\left[ \Phi(L_2,M,0) - \eta(L_2,a) \right] \;.
$

\section{Extraction of HQET hadronic matrix elements and preliminary results}

In the large volume limit, at the first order of the $1/\mb$ 
expansion our main observables are given by
\bes
\mB &=& m_{\rm bare} \,+\,E^{\rm stat}
\,+\, \omega_{\rm kin} \, E^{\rm kin} \,+\,
\omega_{\rm spin} \, E^{\rm spin}\,, 
\\
\mB-m_{\rm B^*} &=& {4\over 3} \omegaspin\,\Espin \;,
\\
\log(a^{3/2}\fB\sqrt{\mB/2}) &=& \log(\zahqet)+ \log(a^{3/2}p^{\rm stat})+ b^{{\rm stat}}_{\rm A} am_{\rm q} 
\nonumber \\&{\phantom{=}}& + \omega_{\rm kin}  p^{\rm kin} +
\omega_{\rm spin}  p^{\rm spin} + \cah{1} p^{\rm A^{(1)}} ,
\label{e:fbmb}
\ees
where $b^{{\rm stat}}_{\rm A}$ is an improvement coefficient.
The HQET energies $E^{\rm stat},E^{\rm kin},E^{\rm spin}$ and the matrix elements 
$p^{\rm stat},p^{\rm A^{(1)}},p^{\rm kin}, p^{\rm spin}$
have been measured on a subset of configuration ensembles 
produced within the CLS effort \cite{CLS} with 
${\rm N_{f}}=2$ flavors of 
${O}(a)$-improved Wilson-Clover fermions. 
We have employed a GEVP analysis in order to control the excited 
states contamination. 
The chiral extrapolations are performed using pion masses
down to 250 MeV (and the coupling $g_{{\rm B}^{*}{\rm B}\pi}$ computed in~\cite{Bulava:2010ej}).
In principle these quantities should be calculated
at various values of the lattice spacing in order
to take the continuum limit.
This continuum extrapolation will be left for future work, instead here we just
compute the observables at one (quite small) value of the lattice spacing
$a=0.07\,\fm$. From our experience in the quenched approximation, 
we do not expect the discretization effects to be visible within 
our present accuracy. More details of this computation can be found 
in~\cite{Blossier:2010vj}. 

\section{Preliminary results and conclusion}
For the b-quark mass, including the $1/\mb$ terms, we find :
\begin{equation}
\left.\mb^{\overline{\rm MS}}(\mb^{\overline{\rm MS}})\right\vert^{{\rm HQET}}_{{\rm N_f}=2}=
4.276(25)_{r_0}(50)_{\rm{stat+renorm}}(?)_a\;\;
\mbox{GeV}\,.
\end{equation}
where the first error comes from the uncertainty on $r_0$, while the second error includes the statistical 
error on $aE^{\rm stat}$, the uncertainty on the chiral extrapolation and the error on the
quark mass renormalization constant.  
At the same order our result for the heavy-light decay constant $\fB$ reads
\be
{\fB}^{\rm HQET}_{\; \rm N_{\rm f}=2}=178(16)(?)_{a}\,\MeV\,,  
\ee
where the first error includes the statistical uncertainty on matrix elements, the systematics coming
from chiral extrapolation and the uncertainty on the physical scale of $r_0$.
In both cases the ``$(?)_a$'' indicates that a continuum limit is not yet performed,
but we expect the discretization effects to be small.
These first results are promising and once we have controlled cut-off effects 
by simulation at several lattice spacings, we plan to extent our project 
to the computation of 
other hadronic quantities like the $B-\bar{B}$ mixing or $B \to \pi$ semileptonic form factors
as well as 
the spectrum of hadrons with a b-flavor.\\

\vspace{-0.5cm}
I am indebted to my colleagues of the ALPHA collaboration, 
in particular I would like to thank 
F.~Bernardoni, B.~Blossier, J.~Bulava, M.~Della~Morte, 
M.~Donnellan, P.~Fritzsch, J.~\mbox{Heitger}, G.~von~Hippel, T.~Mendes, H.~Simma and R.~Sommer
for many stimulating discussions and contributions to this work.



\bibliography{ichep}{}

\begin{thebibliography}{10}

\bibitem{Blossier:2010vj}
ALPHA, {B.~Blossier, J.~Bulava, M.~Della~Morte, M.~Donnellan, P.~Fritzsch,
  N.~Garron, J.~Heitger, G.~von Hippel, B.~Leder, H.~Simma and R.~Sommer},
\newblock PoS {\bf LATTICE2010}, 308 (2010), 1012.1357.

\bibitem{CLS}
{Coordinated Lattice Simulations
  \emph{https://twiki/cern.ch/twiki/bin/view/CLS/WebHome}}.

\bibitem{Aubin:2009yh}
C.~Aubin,
\newblock PoS {\bf LATTICE2009}, 007 (2009), 0909.2686.

\bibitem{DellaMorte:2010ym}
M.~Della~Morte,
\newblock PoS {\bf ICHEP2010}, 364 (2010), 1011.5974.

\bibitem{Gamiz}
E.~G{\'a}miz,
\newblock PoS {\bf ICHEP2010}, 366 (2010).

\bibitem{Sommer:2010ic}
R.~Sommer,
\newblock (2010), 1008.0710.

\bibitem{Blossier:2010vz}
ALPHA, {B.~Blossier , M.~Della Morte , N.~Garron , G.~von Hippel , T.~Mendes ,
  H.~Simma and R.~Sommer},
\newblock JHEP {\bf 05}, 074 (2010), 1004.2661.

\bibitem{DellaMorte:2006cb}
ALPHA, M.~Della~Morte, N.~Garron, M.~Papinutto, and R.~Sommer,
\newblock JHEP {\bf 01}, 007 (2007), hep-ph/0609294.

\bibitem{Blossier:2010mk}
ALPHA, {B.~Blossier , M.~Della Morte , N.~Garron , G.~von Hippel , T.~Mendes ,
  H.~Simma and R.~Sommer},
\newblock JHEP {\bf 12}, 039 (2010), 1006.5816.

\bibitem{Blossier:2010jk}
ALPHA, B.~Blossier, M.~Della~Morte, N.~Garron, and R.~Sommer,
\newblock JHEP {\bf 1006}, 002 (2010), 1001.4783.

\bibitem{Bulava:2010ej}
ALPHA, J.~Bulava, M.~A. Donnellan, and R.~Sommer,
\newblock PoS {\bf LATTICE2010}, 303 (2010), 1011.4393.

\end{thebibliography}
\bibliographystyle{h-physrev}

\end{document}